\documentclass[reprint,aps,prl,superscriptaddress,noeprint]{revtex4-2}
\usepackage{graphicx}
\usepackage{dcolumn}
\usepackage{bm}
\usepackage{amsmath,amssymb,bm}
\usepackage{graphicx}
\usepackage{epstopdf}
\usepackage{latexsym}
\usepackage[normalem]{ulem} 
\usepackage[caption=false]{subfig}
    \usepackage[usenames,dvipsnames]{color}
\usepackage{hyperref}
\usepackage{natbib}
\usepackage{bbold}
\usepackage{comment}
\usepackage{soul}
\usepackage{siunitx}

\hypersetup{
	colorlinks=true,        
}

\begin{document}

\newcommand{\mycomment}[1]{}
\newcommand{\Tr}[1]{\operatorname{Tr}( #1 )}
\newcommand{\ket}[1]{\lvert #1 \rangle}
\newcommand{\bra}[1]{\langle #1 \rvert}
\newcommand{\ketbra}[2]{\ket{#1}\bra{#2}}
\newcommand{\expt}[1]{\langle #1 \rangle}
\renewcommand{\mod}[1]{\lvert #1 \rvert}
\newcommand{\modsq}[1]{\mod{#1}^2}
\newcommand{\partialD}[2]{\frac{\partial #1}{\partial #2}}
\newcommand{\braket}[2]{\langle #1 | #2 \rangle}
\newcommand{\sinc}{\mathrm{sinc}}
\newcommand{\nb}{\mathcal{N}_{\mathrm{b}}}

\newcommand{\warn}[1]{{\color{red}\textbf{* #1 *}}}
\newcommand{\warntoedit}[1]{{\color{blue}\textbf{EDIT: #1 }}}
\newcommand{\warncite}[1]{{\color{green}\textbf{cite #1}}}
\newcommand{\jm}[1]{{\color{magenta} #1 }}

\newcommand{\Rev }[1]{{\color{black}{#1}\normalcolor}} 
\newcommand{\Com}[1]{{\color{red}{#1}\normalcolor}} 

\makeatletter
\def\maketitle{
\@author@finish
\title@column\titleblock@produce
\suppressfloats[t]}
\makeatother

\newcommand{\mytitle}{Mediating gates between polar molecules using microwave-dressed Rydberg atoms }

\title{\mytitle}
\date{\today}

\newcommand{\affA}{Institute for Theoretical Physics, Institute of Physics, University of Amsterdam, Science Park 904, 1098 XH Amsterdam, the Netherlands}
\newcommand{\affB}{QuSoft, Science Park 123, 1098 XG Amsterdam, the Netherlands}
\newcommand{\affCa}{Joint Center for Quantum Information and Computer Science, NIST/University of Maryland, College Park, Maryland 20742, USA}
\newcommand{\affCb}{Joint Quantum Institute, NIST/University of Maryland, College Park, Maryland 20742, USA}

\title{\mytitle}
\date{\today}

\author{Bas Gerritsen}\affiliation{\affA}\affiliation{\affB}

\author{Sean R. Muleady}\affiliation{\affCa}\affiliation{\affCb}

\author{Arghavan Safavi-Naini}\affiliation{\affA}\affiliation{\affB}
\author{Alexey V. Gorshkov}\affiliation{\affCa}\affiliation{\affCb}
\author{Jeremy T. Young}\affiliation{\affA}\affiliation{\affB}
\begin{abstract}
We propose a scheme for mediating many simultaneous, fast entangling gates between pairs of polar molecules using Rydberg atoms. By using microwave drives, the dipolar interactions between Rydberg atoms are nullified. Using the freedom left in the microwave dressing parameters, the Rydberg van der Waals interactions are minimized and the Rydberg-molecule interaction is tuned into resonance, allowing for the mediation of a modified iSWAP gate between molecules.
In the example of mediating gates between $\rm{{}^{23}Na {}^{133}Cs}$ molecules with $\rm {}^{133}Cs$ atoms, the gate is more than two orders of magnitude faster than an equivalent direct molecule-molecule gate.
We model the decay of the dressed Rydberg states and the motion of the atom and molecules and obtain a leading-order estimate of the resulting gate infidelities.
Finally, we show that by detecting the state of the Rydberg atom after a gate, a subset of gate errors can be converted into erasure errors and mitigated by post-selection or quantum error-correction.

\end{abstract}
\maketitle  
Polar molecules are a promising platform for applications in quantum simulation and quantum computation \cite{DeMille2002,cornish2024,Kaufman2021}. These molecules offer both a rich variety of long-lived internal states as well as long-range dipole-dipole interactions.
Simultaneously, Rydberg atoms exhibit even stronger interactions, both dipole-dipole and 
van der Waals, albeit with more susceptibility to dissipation.
Tweezer arrays of neutral atoms have positioned themselves as one of the leading platforms for quantum science \cite{Endres2016,Barredo2016,Browaeys2020,Kaufman2021,Bluvstein2024,Manetsch2024}. In light of recent advances in the capabilities of tweezer arrays of polar molecules \cite{Liu2018,He2020a,Zhang2022,Anderegg2019,Burchesky2021,Bao2023,Holland2023b,
Ruttley2023,Guttridge2023,Park2023,Ruttley2024,Picard2024,Lu2024,Bao2024,Picard2024a, Ruttley2025}, hybrid arrays of both atoms and molecules represent a promising new direction.

By choosing suitable Rydberg states, strong interactions between polar molecules and Rydberg atoms can be realized \cite{zhu2025probing,Guttridge2023,ruttley2026harnessing}.
These interactions between  atoms and molecules can be used for a variety of applications, such as generating entanglement between atoms and polar molecules~\cite{ruttley2026harnessing}, state-insensitive cooling \cite{young2025data} of polar molecules, non-destructive readout of polar molecules \cite{ruttley2026harnessing,kuznetsova2016rydberg,Zhang2022a,Wang2022,Young2026detection} and mediation of entangling gates \cite{Zhang2022a,Wang2022} or interactions \cite{centralspinpaper} between molecules.
We propose a method for mediating a gate between two polar molecules via an intermediary Rydberg atom. In contrast to related prior proposals \cite{Zhang2022a,Wang2022}, we use microwave dressing~\cite{Young2020a,Young2026detection} to minimize unwanted Rydberg-Rydberg interactions, thereby enabling many simultaneous molecular gates.

\begin{figure}[ht]
    \centering
    \includegraphics[width=0.99\linewidth]{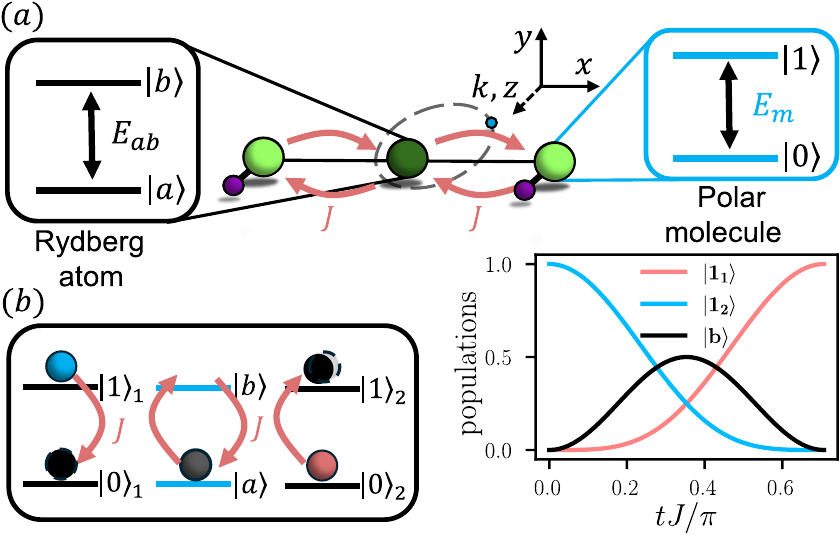}
    \caption{(a) Relevant molecular and Rydberg atom energy levels. When tuned to $E_{ab}=E_m$, the molecules can resonantly exchange excitations with the atom at rate $J$.
    (b) Schematic drawing and plot of state populations under exchange gate dynamics. After gate time $t_G=\pi/(J\sqrt{2})$, the molecules have exchanged excitations while the Rydberg atom is returned to its original state.}
    \label{fig:fig1gate}
\end{figure}

Interactions between molecules are mediated via a pair of Rydberg states $\ket{a},\ket{b}$ which have a transition frequency close to that of the transition between two rotational states $\ket{0},\ket{1}$ of a polar molecule, realizing resonant dipolar flip-flop interactions between atoms and molecules as shown in Fig.~\ref{fig:fig1gate}. 
By positioning a Rydberg atom between each pair of molecules, the resonant dipolar interaction with the atom mediates a modified iSWAP gate~\cite{Yung2003AnEE} between the two molecules, returning the atom to its original state.
In order to simultaneously apply many Rydberg-mediated gates, the Rydberg-molecule interaction needs to be significantly stronger than the other interactions. 
As the transition dipole moments in a Rydberg atom are typically a factor of $\sim 10^3$ larger than the dipole moment of a polar molecule, Rydberg-Rydberg interactions dominate.
To reduce these interactions, we will make use of microwave dressing to engineer dressed Rydberg states which have a negligible dipole-dipole interaction with other Rydberg atoms \cite{Young2020a,Young2026detection}. Furthermore, through careful tuning of the drive parameters, subject to the constraint of nullifying the dipolar Rydberg interactions, the remaining Rydberg van der Waals (vdW) interactions can be minimized as well.
 \begin{figure}
    \centering
    \includegraphics[width=0.99\linewidth]{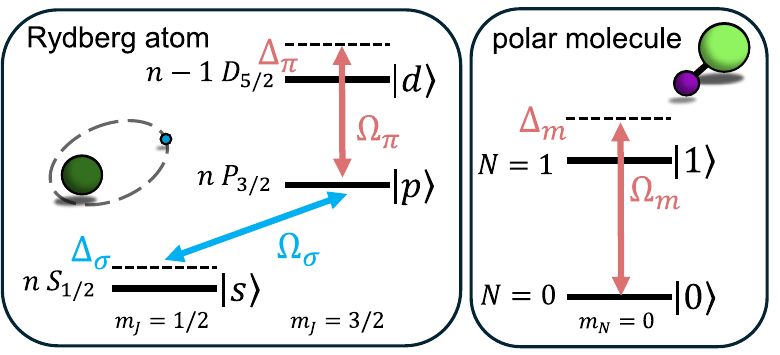}
    \caption{Left: Microwave dressing scheme used to nullify the dipole-dipole interactions between Rydberg atoms. 
    A Rydberg p ($L=1$) state is coupled to a d ($L=2$) state by a $\pi$-polarized microwave field with Rabi frequency $\Omega_\pi$ and detuning $\Delta_\pi$ and a s ($L=0$) state by a $\sigma_+$-polarized field with Rabi frequency $\Omega_\sigma$ and detuning $\Delta_\sigma$. Here, $n$ denotes the principal quantum number.
    Right: The polar molecules are dressed off-resonantly with detuning $\Delta_m$ by the same $\pi$-polarized microwave field used to dress the Rydberg atoms.}
    \label{fig:dressingscheme}
\end{figure}

\emph{Rydberg-molecule interaction}.---We first turn our attention to the interaction between Rydberg atoms and polar molecules.
In the polar molecules, we will consider two rotational states labeled as $\ket{0},\ket{1}$.
Due to the presence of the Rydberg dressing microwave field which is required to nullify dipolar interactions between Rydberg atoms, the molecules will also be driven.
The Rydberg dressing scheme is shown schematically in Fig.~\ref{fig:dressingscheme} and will be explained further in the following section.

In the rotating frame and under the rotating wave approximation, the molecular dressing Hamiltonian is given by
\begin{equation}
    H_{\rm mol}= -\Delta_{m} \ketbra{0}{0} +\Omega_m \ketbra{0}{1} +\Omega_m \ketbra{1}{0}, 
\end{equation}
where the detuning of the drive on the molecule is given by $\Delta_m=\Delta_{\pi} +\omega_R-\omega_m$, $\omega_R$ and $ \omega_m$ are the Rydberg and molecule transition frequencies.
The Rabi frequency is given by $\Omega_m=\Omega_{\pi} \mu_m/ \mu_{\pi}$, while $\Omega_\pi$ and $\Delta_\pi$ are the Rabi frequency and detuning of the microwave field driving the $\pi$ transition in the Rydberg atom and will be discussed further in the next section. 
The transition dipole moments of the molecule and Rydberg atom are given by $\mu_m= \bra{0} d_{0}^m \ket{1}$ and $\mu_\pi= \bra{d} d_0^R \ket{p}$, $\mu_\sigma= \bra{s} d_+^R \ket{p}$ respectively, where $d^{j}_\epsilon=\pmb{{\hat{e}}}_\epsilon \cdot \pmb{d}^{j}$ is the spherical $\epsilon$ component of the dipole operator $\pmb{d}^{j}$ acting on particle $j$.
Here $j=R,m$ label the Rydberg atoms and molecules respectively and $\pmb{\hat{e}}_0=\pmb{z}$, $\pmb{\hat{e}_\pm}=\mp  \frac{1}{\sqrt{2}} (\pmb{x}\pm i \pmb{\hat{y}})$.
Considering $\mu_m$ is much smaller than $\mu_{\pi, \sigma}$, the Rabi frequency on the molecule $\Omega_m$ is likewise lower than on the Rydberg atom. 
To reduce the number of constraints on the Rydberg dressing and avoid requiring very large Rydberg Rabi frequencies, we assume that the molecular drive is both weak and off-resonant.
At the operating point chosen in this manuscript, the corresponding molecular dressed states are approximately the bare $\ket{0},\ket{1}$ states.
In the same rotating frame as the atoms, the energy splitting of the molecules then is $E_{m}\approx\Delta_m$. 

The dipole interaction between polar molecule $i$ and Rydberg atom $R$ in the rotating frame is given by
\begin{equation}
     \begin{split}
        V_{i,R}= &\frac{1-3 \cos^2{\theta_{i,R}}}{4\pi \epsilon_0 r_{i,R}^3}  \mu_\pi \mu_m \ketbra{ 1_ip_R}{0_id_R}+H.c.,
        \end{split}
\label{eqn:V_rydmol}
\end{equation}
where $\theta_{i,R}$ is the angle the displacement vector makes with the quantization axis, $r_{i,R}$ is the distance between molecule $i$ and Rydberg atom $R$.
Using the dressing parameters given in Eq.~(\ref{eqn:stateparams}) for the Rydberg dressed states, the  dressed Rydberg-molecule flip-flop interaction $J$ is proportional to 
\begin{equation}
    J_i\propto \frac{1-3 \cos^2{\theta_{i,R}}}{ r_{i,R}^3} \mu_\pi \mu_m \frac{1-\alpha^2}{\sqrt{1+2\mathcal{M}^2}},
\label{eqn:J_rydmol}
\end{equation}
where $\alpha$ is a Rydberg dressing parameter which is further discussed in the following section and $\mathcal{M}=\mu_\pi/\mu_\sigma$.
In order to obtain a large interaction strength, we need to find a set of Rydberg states with a high value of $\mu_\pi/\sqrt{1+2\mathcal{M}^2}$, requiring both large dipole moments and a favorable ratio between $\mu_\pi$ and $\mu_\sigma$. Furthermore, the mismatch in the relevant transition frequencies of the dressed atoms and molecules should be small in order for the Rydberg-molecule interaction to be tuned to resonance.

If the mismatch between the transition frequencies of the dressed Rydberg atoms and molecules $\Delta E=E_m-E_{ab}=E_m-E_b+E_a$ is much smaller than the interaction strength $J$, a resonant two-molecule gate can be performed using a single Rydberg atom.
At resonance, the Hamiltonian describing the interaction between a single atom and two equidistant molecules is given by
\begin{equation}
    H_G=\sum_{i=1,2} J_{i} \left(\ket{ 1_i a}\bra{0_ib} + \ket{0_i b}\bra{1_i a}\right),
    \label{eqn:H_G}
\end{equation}
where $i$ labels the molecules and the interaction strength $J_i=C_J(\theta_{i,R})/r_{i,R}^3$ is set by the angle-dependent interaction coefficient $C_J(\theta)$ and the molecule-atom distance $r_{i,R}$.
When making two molecules interact with an atom initialized in $\ket{a}$ and $J_1=J_2$, the atom is returned to its original state after a time $t_G=\frac{\pi}{J\sqrt{2}}$, which corresponds to a two-molecule entangling gate.
The result is a modified iSWAP gate~\cite{Yung2003AnEE} which maps the two-molecule states $\ket{00},\ket{10},\ket{01},\ket{11}$ according to 
\begin{equation}
U_{G}=
\begin{pmatrix}
1 & 0 & 0 & 0 \\
0 & 0 & -1 & 0 \\
0 & -1 & 0 & 0 \\
0 & 0 & 0 & -1 
\end{pmatrix}=i R_z^1(\pi/2) ~ R_z^2(\pi/2) ~{\rm{iSWAP}},
\label{eqn:U_gate}
\end{equation}
where $R_z^1(\theta)$ is a z rotation by angle $\theta$ on qubit 1.
This gate swaps excitations between the two molecules and flips the phase of the states that can interact with the atom in state $\ket{a}$. 
The $\ket{11}$ state also couples to the Rydberg atom and undergoes a full $2\pi$ rotation during the gate, thus acquiring a $-1$ phase factor.
As $U_G$ is equivalent to the iSWAP gate up to single-qubit rotations, it can be used for universal quantum computation if combined with arbitrary single-qubit gates~\cite{echternach2001universal}.

\emph{Rydberg microwave dressing}.---For the mediation of simultaneous gates between molecules, it is important for the dipolar interactions between Rydberg atoms to be minimized.
In order to nullify the interactions between atoms, we consider the dressing scheme shown in Fig.~\ref{fig:dressingscheme}. Although we specifically study the case where a Rydberg p state ($L=1$) is coupled to a d ($L=2$) state by a $\pi$-polarized microwave field and a s ($L=0$) state by a $\sigma_+$-polarized field, the scheme is general and only requires one microwave field to drive a $\pi$ transition and the other to drive a $\sigma_+$ transition.
The dressing scheme could be extended by including more drives, which would provide a greater degree of tunability.

In the rotating frame and under the rotating wave approximation, the Rydberg microwave dressing Hamiltonian is given by
\begin{equation}
\begin{split}
 H_{\rm mw} = & - \Delta_\pi \ketbra{d}{d} + \Omega_\pi \ketbra{p}{d} +\Omega_\pi \ketbra{d}{p} \\
   & - \Delta_\sigma \ketbra{s}{s} + \Omega_\sigma \ketbra{p}{s} +\Omega_\sigma \ketbra{s}{p}. 
   \end{split}
\label{eqn:MW_Hamiltonian}
\end{equation}
The three states in our dressing scheme all have different orbital angular momenta and are connected through the central $\ket{p}$ state by dipole transitions.
Therefore the dressed states of $H_{\rm mw}$ also experience dipolar interactions.
In the rotating frame, atoms i and j interact via the dipole-dipole interaction
\begin{equation}
    \begin{split}
        V_{dd}^{ij}= &\frac{1-3 \cos^2{\theta_{i,j}}}{4\pi \epsilon_0 r_{i,j}^3} ( \mu_\pi^2 \ketbra{p_id_j}{d_ip_j} \\
        & -\mu_\sigma^2/2 \ketbra{p_i s_j}{s_i p_j} ) +H.c.,
        \end{split}
\label{eqn:Vddrydryd}
\end{equation}
where $\theta_{i,j}$ is the angle the displacement vector makes with the quantization axis and $r_{i,j}$ is the distance between the atoms.

Next, we discuss how to engineer the dressed-state parameters to nullify the dipole-dipole interactions between atoms.
By applying two microwave drives with different polarizations, we can make use of a sign difference between the dipolar interactions caused by each drive in Eq.~(\ref{eqn:Vddrydryd}) and tune the dressed Rydberg states so all dipolar interactions between the atoms destructively interfere.
We consider two dressed states $\ket{a},\ket{b}$, which are eigenstates of $H_{\rm mw}$ and have real coefficients
\begin{equation}
    \begin{split}
        \ket{a} & =  a_p \ket{p} + a_\pi \ket{d} + a_\sigma \ket{s},\\
        \ket{b} & = b_p \ket{p} + b_\pi \ket{d} + b_\sigma \ket{s}.
    \end{split}
    \label{eqn:state3}
\end{equation}
The relevant dipole-dipole interactions between the pair states $\ket{aa},\ket{bb},\ket{ab}, \ket{ba}$ are proportional to 
\begin{equation}
    \begin{aligned}
        V_{aa}=&~\bra{aa}V_{dd}\ket{aa} \propto  a_p^2 \left(a_\sigma^2 \mu_\sigma^2-2 a_\pi^2 \mu_\pi^2 \right),\\
        V_{bb}=&~ \bra{bb}V_{dd}\ket{bb} \propto b_p^2 \left(b_\sigma^2 \mu_\sigma^2-2 b_\pi^2 \mu_\pi^2 \right),\\
         V_{ab} =&~ \bra{ab}V_{dd}\ket{ab} \propto a_p b_p \left( a_\sigma b_\sigma \mu_\sigma^2 -2a_\pi b_\pi \mu_\pi^2\right), \\
          \tilde{V}_{ab}=&~ \bra{ab}V_{dd}\ket{ba} \propto\left (a_\sigma^2 b_p^2 +a_p^2 b_\sigma^2\right)\mu_\sigma^2 \\ 
          & ~- 2\left(a_\pi^2 b_p^2 +a_p^2 b_\pi^2 \right)\mu_\pi^2. 
    \end{aligned}
\label{eqn:V_dd_constraints}
\end{equation}
We require the dressed states $\ket{a},\ket{b}$ to be normalized, orthogonal eigenstates of the dressing Hamiltonian given in Eq.~(\ref{eqn:MW_Hamiltonian}).
From Eqs.~(\ref{eqn:state3},~\ref{eqn:V_dd_constraints}) and the dressed-state constraints, it can be shown that if $V_{aa},V_{ab}$ are nullified, the other dipolar interactions will also be nullified.

In order to nullify the dipole-dipole interactions using the considered microwave drive, five constraints need to be satisfied while there are six state parameters. 
After imposing normalisation, orthogonality, the eigenstate conditions and the nullification of $V_{aa},V_{ab}$, a single free parameter remains.
We label this parameter as $\alpha$, which is related to the ratio between the detunings and Rabi frequencies.
The microwave drive Hamiltonian $H_{\rm mw}$ can be freely rescaled by a constant; this freedom is absorbed via $\Omega_\pi$.
These two degrees of freedom can be used to both tune dressed-state energies to bring the Rydberg-molecule interaction close to resonance and also minimize the Rydberg-Rydberg vdW interactions.
The definition of $\alpha$, the full expressions for the required dressing parameters and a discussion of the allowed parameter ranges are presented in the End Matter, while the role of the third dressed state is discussed in the End Matter.

\emph{vdW interactions}.---For the mediation of entangling molecule-molecule gates, not only the direct dipolar interactions, but also the Rydberg-Rydberg vdW interactions need to be minimized.
Although the dipolar interactions have been mitigated through the microwave dressing, the vdW interactions between neighboring Rydberg atoms can still be significantly stronger than the dipolar Rydberg-molecule interactions at distances of a few $\si{\micro \metre}$.
Due to the presence of the microwave drive, the energy splitting and off-resonant dipolar couplings between the dressed states are modified when the drive parameters are changed, thus allowing the vdW interactions to be tuned. 
When a pair of $\ket{a}$ and $\ket{b}$ states (e.g. $\ket{aa}$) becomes degenerate with another pair of Rydberg states, a two-atom resonance occurs and the corresponding vdW interaction flips sign.
Because many such resonances contribute to the vdW coefficients, there are zero-crossings of the vdW coefficients. 
Therefore, using the two tunable parameters, two distinct vdW terms can in principle be nullified completely. 
Enforcing resonance introduces one additional constraint, whereas the dressing provides only two tunable parameters. 
The four interactions in Eq.~(\ref{eqn:C6}) can therefore generally be reduced but not simultaneously nullified.
Because the dipole interactions involved in the vdW processes have multiple different angular dependences, minimizing the vdW interactions is only possible for a single value of $\theta_{i,j}$; operating at a different $\theta_{i,j}$ would require a nontrivial change of drive parameters to minimize the vdW interactions again.
This is in contrast to the nullification of the dipole-dipole interactions in Eq.~(\ref{eqn:Vddrydryd}), which is valid at all $\theta_{i,j}$.

\begin{figure}[ht]
    \centering
    \includegraphics[width=1.\linewidth]{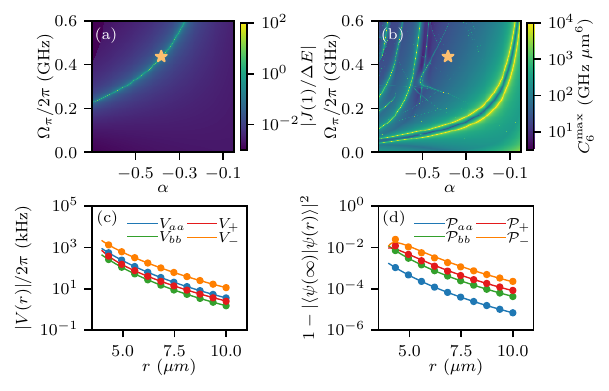}
    \caption{(a) $|J(1)/\Delta E|$ at $\theta=\pi/2$ as a function of dressing parameters, where $J(1)$ is the Rydberg-molecule interaction strength at $r_{\rm Rm}=1$ $\si{\um}$ and $\Delta E$ is the mismatch in the Rydberg and molecule transition frequencies. The yellow star denotes the values of $(\alpha,\Omega_\pi)$ used in (c) and (d).
    (b) The largest of the four relevant vdW coefficients $C_6^{\rm{max}}=\max{ \left(|C_6^{aa}|,|C_6^{bb}|,|C_6^{ab}|,|\tilde{C}_6^{ab}| \right)}$ at $\theta=\pi/2$ as a function of the dressing parameters.  Here a small $C_6^{\rm{max}}$ indicates that all the relevant vdW coefficients are small.
    (c) Minimized Rydberg vdW potentials (points) at $\theta=\pi/2$ and fits (solid lines) with $V_{ij}=-C^{ij}_6/r^6-C^{ij}_9/r^9-C^{ij}_{12}/r^{12}$, where $\mathbf{C}^{ij}=(C^{ij}_6,C^{ij}_9,C^{ij}_{12})$ $\mathbf{C}^{aa}/2\pi$~=~$(-3.58 ~ \qty{}{\giga\hertz\micro\metre}^6,4.58 ~ \qty{}{\giga\hertz\micro\metre}^9, -2.15 ~ \qty{}{\giga\hertz\micro\metre}^{12})$,
    $\mathbf{C}^{bb}/2\pi$~=~$(1.48 ~ \qty{}{\giga\hertz\micro\metre}^6,13.7 ~ \qty{}{\giga\hertz\micro\metre}^9, 98.3 ~ \qty{}{\giga\hertz\micro\metre}^{12})$,
    $\mathbf{C}^{+}/2\pi$~=~$(2.63 ~ \qty{}{\giga\hertz\micro\metre}^6,122~ \qty{}{\giga\hertz\micro\metre}^9, 8.56 \times 10^3 ~ \qty{}{\giga\hertz\micro\metre}^{12})$ and $\mathbf{C}^{-}/2\pi$~=~$(11.7 ~ \qty{}{\giga\hertz\micro\metre}^6,-476~ \qty{}{\giga\hertz\micro\metre}^9, 17.7 \times 10^4 ~ \qty{}{\giga\hertz\micro\metre}^{12})$.
    We define $\ket{\pm}=\left(\ket{ab} \pm\ket{ba}\right)/\sqrt{2}$.
    (d) Admixture error of the four relevant two-atom states at $\theta=\pi/2$ and corresponding fits (solid lines) with $\mathcal{P}_{ij}=P^{ij}_6/r^6+P^{ij}_{12}/r^{12}$, where $\mathbf{P}^{ij}=(P_6^{ij},P_{12}^{ij})$, $\mathbf{P}^{aa}= (6.79 ~ \qty{}{\micro\metre}^6, -104~ \qty{}{\micro\metre}^{12})$,
    $\mathbf{P}^{bb}= (42.0 ~ \qty{}{\micro\metre}^6, -2.02 \times 10^4~ \qty{}{\micro\metre}^{12})$,  $\mathbf{P}^{+}= (79.8 ~ \qty{}{\micro\metre}^6, -1.70 \times 10^5~ \qty{}{\micro\metre}^{12})$ and  $\mathbf{P}^{-}= (215 ~ \qty{}{\micro\metre}^6, -6.79 \times 10^5~ \qty{}{\micro\metre}^{12})$.}
    \label{fig:ryd-mol_resonance}
\end{figure}

The relevant vdW interactions are calculated using the same procedure outlined in~\cite{Young2020a} and take the form
\begin{equation}
    \begin{split}
        V_{\rm vdW} &=-\frac{C_6^{aa}(\theta)}{r^6} \ket{aa}\bra{aa} -   \frac{C_6^{bb}(\theta)}{r^6} \ket{bb}\bra{bb}\\
        &-\frac{C_6^{ab}(\theta)}{r^6} \left( \ket{ab}\bra{ab} +\ket{ba}\bra{ba} \right)\\
        &-\frac{\tilde{C}_6^{ ab}(\theta)}{r^6} \left( \ket{ab}\bra{ba} + \rm{H.c.} \right),
    \end{split}
\label{eqn:C6}
\end{equation}
where $C_6^{aa},C_6^{bb},C_6^{ab}$ are the vdW coefficients of the diagonal interactions for $\ket{a},\ket{b} ,\ket{ab}$ respectively and $\tilde{C}_6^{ab}$ is the coefficient for the off-diagonal $\ket{ab}\leftrightarrow \ket{ba}$ interaction. 
We illustrate the tunability of the vdW interactions using the example of Cs Rydberg atoms in Fig.~\ref{fig:ryd-mol_resonance}. Although there are four relevant vdW interactions, we only show the $C_6$ coefficient for each pair of values of $(\alpha_,\Omega_\pi)$ as the largest interaction will typically dominate the dynamics.
When tuning $C_6$ coefficients close to zero, higher order terms in the perturbative expansion can become important. 
We capture these terms by fitting the interaction potential between two Rydberg pair states, e.g., $\ket{aa}$ with $V^{aa}=-C^{aa}_6/r^6-C^{aa}_9/r^9-C^{aa}_{12}/r^{12}$.

\emph{NaCs and Cs}.---As an example of the parameters needed to realize the mediation scheme described above, we consider NaCs molecules and Cs Rydberg atoms.
NaCs has a rotational constant of $B_\nu=2 \pi \times 1.74$~GHz~\cite{wang2024dual} and a permanent dipole moment of $4.6$~D~\cite{PhysRevA.88.032709}.
We will focus on the $\ket{N=0,m_N=0} \leftrightarrow \ket{N=1,m_N=0}$ rotational transition in NaCs and therefore target a transition frequency of $\omega_m= 2 B_\nu= 2\pi \times 3.48$~GHz.
Choosing a $n~p_{3/2} \leftrightarrow (n-1)d_{5/2}$ transition in Cs, the closest bare transition frequency  match to NaCs occurs at $n=59$.
However, choosing a different principal quantum number can be preferred.
The transition frequency mismatch $\Delta E$ and the associated values of $(\alpha,\Omega_\pi)$ at which the Rydberg-molecule interaction becomes resonant depend on $n$.
In order to shift the $\Delta E=0$ resonance to an $\alpha,\Omega_\pi$ parameter regime where the Rydberg vdW interactions can be minimized more effectively, we choose $n=57$.
Labeling states by $\ket{n,L,J,m_J}$, the Rydberg states used in the dressing scheme are $\ket{s}= \ket{57,0,1/2,1/2}$, $\ket{p}= \ket{57,1,3/2,3/2}$ and $\ket{d}= \ket{56,2,5/2,3/2}$.
Fig.~\ref{fig:dressingscheme} shows a schematic depiction of the dressing scheme.
These states were chosen to maximize the Rydberg-molecule interaction strength given in Eq.~(\ref{eqn:V_rydmol}) by having a favorable $\mu_\pi /\mu_\sigma$ ratio.
The corresponding transition dipole moments are $\mu_\pi=4329$ D, $\mu_\sigma= 4687$ D ~\cite{ARC2017}, thus $\mathcal{M}=0.924$.
For the molecule, the transition dipole moment is $\mu_m=4.6/\sqrt{3}$ D.

Throughout the rest of this manuscript, we set $\theta=\pi/2$ in order to suppress dipolar atom–molecule interactions that change the $m_N$ projection of the molecule.
Analyzing the $\alpha,\Omega_\pi$ dependence of the vdW coefficients shown in Fig.~\ref{fig:ryd-mol_resonance}, we observe several regions where the vdW interactions are minimized. 
The resonance around $\Delta E=0$ in panel (a) overlaps with a region where the $C_6$ coefficients are relatively small at higher $\Omega_\pi$ in panel (b), making this a suitable parameter regime to operate in.
As an example, we pick the point $\bigstar$ at $\Omega_\pi=2\pi \times 0.44$ GHz$,\alpha=-0.38$ on the resonance as indicated by the yellow star in Fig.~\ref{fig:ryd-mol_resonance}.
At this point $|\Delta_m/\Omega_m|\approx3000$, making the weak, off-resonant dressing assumption valid.
Around point $\bigstar$, the Rydberg-molecule interaction coefficient is $C_J(\theta=\pi/2) =2\pi \times 960$ kHz $\si{\um}^{3}$, corresponding to a gate time of $t_G=0.37$ $\si{\us}$ at $r_{\rm Rm} =1~\si{\um}$, while the decay rates of the dressed states $\ket{a},\ket{b}$ at $293 $ K are $\Gamma_{a,b}/2\pi\approx 1.3,1.7$ kHz~\cite{ARC2017}. 
To make the comparison to a direct molecule-molecule iSWAP gate, we set the molecule-molecule distance to $r_{\rm mm}=1~\si{\um}$ as a best-case scenario and set $\theta=\pi/2$. 
With a transition dipole moment of $\mu_m=4.6/\sqrt{3}$ D, the corresponding gate time is given by $t_G^{\rm iSWAP}=\pi/(2J_{\rm mm})\approx230~\si{\us}$ and is thus more than two orders of magnitude slower than the mediated gate.
Conversely, the error introduced by the direct interaction between molecules (separated by $r_{mm}=2r_{\rm Rm}$) during a mediated gate is negligible.

\emph{Gate errors.}---The decay of the dressed Rydberg atom during the gate, fluctuations of the interaction strength due to (thermal) motion and undesired spin-motion entanglement caused by the dipole forces between the molecules and atom can limit gate performance.
We obtain a leading-order estimate of the associated gate infidelity by modeling the quantized motion of the atom and molecules in 1D and calculating the infidelity caused by motion along each of the three orthogonal axes separately.
These axes are defined according to panel (a) of Fig.~\ref{fig:fig1gate}.
We assume a radial (perpendicular to the tweezer k-vector) trapping frequency of $\omega_{x,y}=2\pi \times 300$ kHz and an axial trapping frequency of $\omega_{z}=2\pi \times 60$ kHz for both the atoms and molecules.
Furthermore, we assume the atom tweezer is switched off during the gate in order to avoid potential issues caused by state-dependent trapping or anti-trapping of the Rydberg state.
The resulting free expansion of the atom wave packet is included in our simulations. 

\begin{figure}
    \centering
    \includegraphics[scale=1]{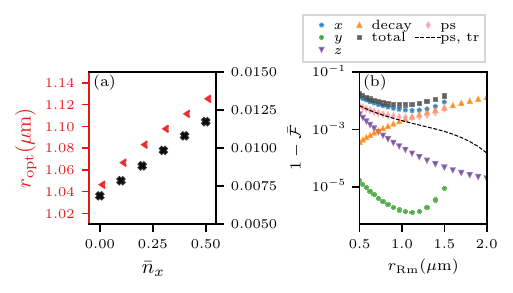}
    \caption{
    (a) Optimal atom-molecule spacing and associated minimum gate average infidelity (without post-selection) at $\theta=\pi/2$ as a function of the thermal excitation number for all particles $\bar{n}_x$. All particles are initialized in $\bar{n}_{y,z}=0$. Due to the increased positional spread of the wave packet of the atom and molecules, it is beneficial to increase the spacing for higher $\bar{n}_x$.
    (b) Average gate infidelity from motion along the $(x,y,z)$ directions and Rydberg decay at $\theta=\pi/2$ as a function of the atom-molecule spacing $r_{\rm Rm}$.
    The total infidelity is a leading-order estimate given by the sum of the $x,y,z$ and decay infidelities.
    The ``ps'' infidelity corresponds to the gate infidelity after post-selection by measuring the state of the atom. 
    The dashed line labeled as ``ps-tr'' corresponds to the post-selected infidelity for a trapped atom.
    All particles are initialized in the motional ground state.}
    \label{fig:errors}
\end{figure}

Using the above parameters and initializing the particles in the motional ground state, the estimated gate infidelity caused by motion and Rydberg decay is $1-\mathcal{F}\sim 10^{-2}$ at an atom-molecule spacing of $r_{\rm Rm}=1.06$ $\si{\um}$, with corresponding gate time $t_G=0.44$  $\si{\us}$.
As shown in panel (a) of Fig.~\ref{fig:errors}, due to the increased spread in the position of the atoms and molecules, the optimal atom-molecule spacing increases with the mean excitation number $\bar{n}_x$ along the $x$-axis, which connects the particles.

Because the Rydberg atom should return to $\ket{a}$ after an ideal gate, a subset of gates which failed due to decoherence caused by Rydberg decay or motion can be detected by measuring the state of the atom. 
These detectable gate errors can be converted into flagged erasure errors and can be reduced using post-selection or quantum error correction. 
This error-mitigation requires high readout fidelities and negligible repopulation of the $\ket{a},\ket{b}$ manifold after the atom leaves this manifold.
Because erasure errors occur at known locations, they can be more easily mitigated by quantum error-correcting codes than errors occuring at unknown locations \cite{grassl1997codes,gottesman1997stabilizer}, allowing a higher physical error rate at the fault-tolerance threshold \cite{wu2022erasure}.
We quantify the average gate infidelity after post-selection by the points labeled ``ps'' in Fig.~\ref{fig:errors}(b), from which it is clear that the reported gate fidelity can be improved by discarding runs where the atom is not detected in $\ket{a}$.  The corresponding fidelity after post-selection for a trapped atom is given by the dashed line labeled as ``ps-tr'' in Fig.~\ref{fig:errors}(b).
The average post-selection rejection rate and more information on the gate error calculations are given in the End Matter.

\emph{Conclusion and outlook.}---We have proposed a scheme for simultaneously mediating many fast entangling gates between polar molecules using microwave-dressed Rydberg atoms.
The performance of the mediated gate proposed here could be further improved by applying an optimal-control scheme to the mediated gate~\cite{jandura2022time,fu2022high,theis2016high}.
Furthermore, as the Rydberg atom should always return to $\ket{a}$ after a gate, some errors caused by decoherence of the Rydberg atom can be detected by measuring the state of the atom and mitigated by subsequent post-selection or erasure conversion and error-correction. 
\\
\\
\emph{Note added.}---During completion of this work we became
aware of a relevant work on quantum logic control and entanglement in hybrid atom-molecule arrays \cite{zhang2026quantum}.
\\
\begin{acknowledgments}
\emph{Acknowledgments.}---We thank Kang-Kuen Ni for insightful discussions and feedback on the manuscript.
S.R.M. is supported by the
NSF QLCI award OMA-2120757. A.V.G.~was supported in part by AFOSR MURI, ONR MURI, the DoE ASCR Quantum Testbed Pathfinder program (awards No.~DE-SC0019040 and No.~DE-SC0024220), NSF QLCI, NSF STAQ program, ARL (W911NF-24-2-0107), and NQVL:QSTD:Design:FTL. A.V.G.~also acknowledges support from the U.S.~Department of Energy, Office of Science, National Quantum Information Science Research Centers, Quantum Systems Accelerator (QSA) and from the U.S.~Department of Energy, Office of Science, Accelerated Research in Quantum Computing, Fundamental Algorithmic Research toward Quantum Utility (FAR-Qu).
A.S.N. and B.G. were supported by the Dutch Research Council (NWO/OCW) as part of the Quantum Software Consortium (project number 024.003.037), QDNL (project number NGF.1582.22.030) and ENW-XL grant (project number OCENW.XL21.XL21.122). J.T.Y.~was supported by the NWO Talent Programme (project number VI.Veni.222.312), which is (partly) financed by the Dutch Research Council (NWO).
\end{acknowledgments}
\bibliographystyle{apsrev4-1}
\bibliography{sample,MoleculeRydberg}

@article{Park2023,
archivePrefix = {arXiv},
arxivId = {2306.07264},
author = {Park, Annie J. and Picard, Lewis R.B. and Patenotte, Gabriel E. and Zhang, Jessie T. and Rosenband, Till and Ni, Kang Kuen},
doi = {10.1103/PhysRevLett.131.183401},
eprint = {2306.07264},
issn = {10797114},
journal = {Phys. Rev. Lett.},
number = {18},
pages = {183401},
pmid = {37977633},
publisher = {American Physical Society},
title = {{Extended Rotational Coherence of Polar Molecules in an Elliptically Polarized Trap}},
url = {https://doi.org/10.1103/PhysRevLett.131.183401},
volume = {131},
year = {2023}
}

@article{He2020a,
author = {He, Xiaodong and Wang, Kunpeng and Zhuang, Jun and Xu, Peng and Gao, Xiang and Guo, Ruijun and Sheng, Cheng and Liu, Min and Wang, Jin and Li, Jiaming and Shlyapnikov, G. V. and Zhan, Mingsheng},
doi = {10.1126/science.aba7468},
issn = {0036-8075},
journal = {Science},
month = {oct},
number = {6514},
pages = {331--335},
pmid = {32972992},
title = {{Coherently forming a single molecule in an optical trap}},
url = {https://www.science.org/doi/10.1126/science.aba7468},
volume = {370},
year = {2020}
}

@article{Ruttley2024,
archivePrefix = {arXiv},
arxivId = {2401.13593},
author = {Ruttley, Daniel K. and Guttridge, Alexander and Hepworth, Tom R. and Cornish, Simon L.},
doi = {10.1103/PRXQuantum.5.020333},
eprint = {2401.13593},
issn = {2691-3399},
journal = {PRX Quantum},
month = {may},
number = {2},
pages = {020333},
title = {{Enhanced Quantum Control of Individual Ultracold Molecules Using Optical Tweezer Arrays}},
url = {http://arxiv.org/abs/2401.13593 https://link.aps.org/doi/10.1103/PRXQuantum.5.020333},
volume = {5},
year = {2024}
}

@article{Burchesky2021,
archivePrefix = {arXiv},
arxivId = {2105.15199},
author = {Burchesky, Sean and Anderegg, Lo{\"{i}}c and Bao, Yicheng and Yu, Scarlett S. and Chae, Eunmi and Ketterle, Wolfgang and Ni, Kang Kuen and Doyle, John M.},
doi = {10.1103/PhysRevLett.127.123202},
eprint = {2105.15199},
issn = {10797114},
journal = {Phys. Rev. Lett.},
number = {12},
pages = {123202},
pmid = {34597100},
publisher = {American Physical Society},
title = {{Rotational Coherence Times of Polar Molecules in Optical Tweezers}},
url = {https://doi.org/10.1103/PhysRevLett.127.123202},
volume = {127},
year = {2021}
}

@article{Young2020a,
archivePrefix = {arXiv},
arxivId = {2006.02486},
author = {Young, Jeremy T. and Bienias, Przemyslaw and Belyansky, Ron and Kaufman, Adam M. and Gorshkov, Alexey V.},
doi = {10.1103/physrevlett.127.120501},
eprint = {2006.02486},
issn = {0031-9007},
title = {{Asymmetric blockade and multi-qubit gates via dipole-dipole interactions}},
url = {http://arxiv.org/abs/2006.02486},
year = {2020}
}

@article{Liu2018,
archivePrefix = {arXiv},
arxivId = {1804.04752},
author = {Liu, L. R. and Hood, J. D. and Yu, Y. and Zhang, J. T. and Hutzler, N. R. and Rosenband, T. and Ni, K. K.},
doi = {10.1126/science.aar7797},
eprint = {1804.04752},
issn = {10959203},
journal = {Science},
number = {6391},
pages = {900--903},
pmid = {29650700},
title = {{Building one molecule from a reservoir of two atoms}},
volume = {360},
year = {2018}
}

@article{Bao2024,
archivePrefix = {arXiv},
arxivId = {2309.08706},
author = {Bao, Yicheng and Yu, Scarlett S. and You, Jiaqi and Anderegg, Lo{\"{i}}c and Chae, Eunmi and Ketterle, Wolfgang and Ni, Kang-Kuen and Doyle, John M.},
doi = {10.1103/PhysRevX.14.031002},
eprint = {2309.08706},
issn = {2160-3308},
journal = {Phys. Rev. X},
month = {jul},
number = {3},
pages = {031002},
title = {{Raman Sideband Cooling of Molecules in an Optical Tweezer Array to the 3D Motional Ground State}},
url = {http://arxiv.org/abs/2309.08706 https://link.aps.org/doi/10.1103/PhysRevX.14.031002},
volume = {14},
year = {2024}
}

@article{Ruttley2023,
archivePrefix = {arXiv},
arxivId = {2302.07296},
author = {Ruttley, Daniel K. and Guttridge, Alexander and Spence, Stefan and Bird, Robert C. and {Le Sueur}, C. Ruth and Hutson, Jeremy M. and Cornish, Simon L.},
doi = {10.1103/PhysRevLett.130.223401},
eprint = {2302.07296},
issn = {10797114},
journal = {Phys. Rev. Lett.},
number = {22},
pages = {223401},
pmid = {37327422},
publisher = {American Physical Society},
title = {{Formation of Ultracold Molecules by Merging Optical Tweezers}},
url = {https://doi.org/10.1103/PhysRevLett.130.223401},
volume = {130},
year = {2023}
}

@article{Endres2016,
author = {Endres, Manuel and Bernien, Hannes and Keesling, Alexander and Levine, Harry and Anschuetz, Eric R and Krajenbrink, Alexandre and Senko, Crystal and Vuletic, Vladan and Greiner, Markus and Lukin, Mikhail D},
doi = {10.1126/science.aah3752},
issn = {0036-8075},
journal = {Science},
month = {nov},
number = {6315},
pages = {1024--1027},
title = {{Atom-by-atom assembly of defect-free one-dimensional cold atom arrays}},
url = {http://www.sciencemag.org/lookup/doi/10.1126/science.aah3752 https://www.science.org/doi/10.1126/science.aah3752},
volume = {354},
year = {2016}
}

@article{Aldegunde2017,
archivePrefix = {arXiv},
arxivId = {1708.05734},
author = {Aldegunde, Jesus and Hutson, Jeremy M.},
doi = {10.1103/PhysRevA.96.042506},
eprint = {1708.05734},
issn = {2469-9926},
journal = {Phys. Rev. A},
month = {oct},
number = {4},
pages = {042506},
title = {{Hyperfine structure of alkali-metal diatomic molecules}},
url = {https://link.aps.org/doi/10.1103/PhysRevA.96.042506},
volume = {96},
year = {2017}
}

@article{Browaeys2020,
archivePrefix = {arXiv},
arxivId = {2002.07413},
author = {Browaeys, Antoine and Lahaye, Thierry},
doi = {10.1038/s41567-019-0733-z},
eprint = {2002.07413},
issn = {1745-2473},
journal = {Nat. Phys.},
month = {feb},
number = {2},
pages = {132--142},
publisher = {Springer US},
title = {{Many-body physics with individually controlled Rydberg atoms}},
url = {http://dx.doi.org/10.1038/s41567-019-0733-z http://www.nature.com/articles/s41567-019-0733-z},
volume = {16},
year = {2020}
}

@article{Zhang2022a,
archivePrefix = {arXiv},
arxivId = {2204.04276},
author = {Zhang, Chi and Tarbutt, M.R.},
doi = {10.1103/PRXQuantum.3.030340},
eprint = {2204.04276},
issn = {2691-3399},
journal = {PRX Quantum},
month = {sep},
number = {3},
pages = {030340},
publisher = {American Physical Society},
title = {{Quantum Computation in a Hybrid Array of Molecules and Rydberg Atoms}},
url = {https://doi.org/10.1103/PRXQuantum.3.030340 https://link.aps.org/doi/10.1103/PRXQuantum.3.030340},
volume = {3},
year = {2022}
}

@article{Bao2023,
archivePrefix = {arXiv},
arxivId = {2211.09780},
author = {Bao, Yicheng and Yu, Scarlett S. and Anderegg, Lo{\"{i}}c and Chae, Eunmi and Ketterle, Wolfgang and Ni, Kang Kuen and Doyle, John M.},
doi = {10.1126/science.adf8999},
eprint = {2211.09780},
issn = {10959203},
journal = {Science},
number = {6675},
pages = {1138--1144},
pmid = {38060651},
title = {{Dipolar spin-exchange and entanglement between molecules in an optical tweezer array}},
volume = {382},
year = {2023}
}

@article{Lu2024,
archivePrefix = {arXiv},
arxivId = {2306.02455},
author = {Lu, Yukai and Li, Samuel J. and Holland, Connor M. and Cheuk, Lawrence W.},
doi = {10.1038/s41567-023-02346-3},
eprint = {2306.02455},
issn = {1745-2473},
journal = {Nat. Phys.},
month = {mar},
number = {3},
pages = {389--394},
publisher = {Springer US},
title = {{Raman sideband cooling of molecules in an optical tweezer array}},
url = {https://www.nature.com/articles/s41567-023-02346-3},
volume = {20},
year = {2024}
}

@article{Ruttley2025,
archivePrefix = {arXiv},
arxivId = {2408.14904},
author = {Ruttley, Daniel K. and Hepworth, Tom R. and Guttridge, Alexander and Cornish, Simon L.},
doi = {10.1038/s41586-024-08365-1},
eprint = {2408.14904},
isbn = {4158602408},
issn = {0028-0836},
journal = {Nature (London)},
month = {jan},
number = {8047},
pages = {827--832},
pmid = {39814895},
publisher = {Springer US},
title = {{Long-lived entanglement of molecules in magic-wavelength optical tweezers}},
url = {http://arxiv.org/abs/2408.14904 https://www.nature.com/articles/s41586-024-08365-1},
volume = {637},
year = {2025}
}

@article{Holland2023b,
archivePrefix = {arXiv},
arxivId = {2210.06309},
author = {Holland, Connor M. and Lu, Yukai and Cheuk, Lawrence W.},
doi = {10.1126/science.adf4272},
eprint = {2210.06309},
issn = {10959203},
journal = {Science},
number = {6675},
pages = {1143--1147},
pmid = {38060644},
title = {{On-demand entanglement of molecules in a reconfigurable optical tweezer array}},
volume = {382},
year = {2023}
}

@article{Zhang2022,
archivePrefix = {arXiv},
arxivId = {2112.00991},
author = {Zhang, Jessie T. and Picard, Lewis R B and Cairncross, William B. and Wang, Kenneth and Yu, Yichao and Fang, Fang and Ni, Kang-Kuen},
doi = {10.1088/2058-9565/ac676c},
eprint = {2112.00991},
issn = {2058-9565},
journal = {Quantum Sci. Technol.},
month = {jul},
number = {3},
pages = {035006},
publisher = {IOP Publishing},
title = {{An optical tweezer array of ground-state polar molecules}},
url = {https://iopscience.iop.org/article/10.1088/2058-9565/ac676c},
volume = {7},
year = {2022}
}

@article{Barredo2016,
archivePrefix = {arXiv},
arxivId = {1607.03042},
author = {Barredo, Daniel and de L{\'{e}}s{\'{e}}leuc, Sylvain and Lienhard, Vincent and Lahaye, Thierry and Browaeys, Antoine},
doi = {10.1126/science.aah3778},
eprint = {1607.03042},
issn = {0036-8075},
journal = {Science},
month = {nov},
number = {6315},
pages = {1021--1023},
title = {{An atom-by-atom assembler of defect-free arbitrary two-dimensional atomic arrays}},
url = {http://arxiv.org/abs/1607.03042 http://www.sciencemag.org/lookup/doi/10.1126/science.aah3778},
volume = {354},
year = {2016}
}

@article{Bluvstein2024,
archivePrefix = {arXiv},
arxivId = {2312.03982},
author = {Bluvstein, Dolev and Evered, Simon J. and Geim, Alexandra A. and Li, Sophie H. and Zhou, Hengyun and Manovitz, Tom and Ebadi, Sepehr and Cain, Madelyn and Kalinowski, Marcin and Hangleiter, Dominik and {Bonilla Ataides}, J. Pablo and Maskara, Nishad and Cong, Iris and Gao, Xun and {Sales Rodriguez}, Pedro and Karolyshyn, Thomas and Semeghini, Giulia and Gullans, Michael J. and Greiner, Markus and Vuleti{\'{c}}, Vladan and Lukin, Mikhail D.},
doi = {10.1038/s41586-023-06927-3},
eprint = {2312.03982},
issn = {0028-0836},
journal = {Nature (London)},
month = {dec},
number = {7997},
pages = {58--65},
pmid = {38056497},
publisher = {Springer US},
title = {{Logical quantum processor based on reconfigurable atom arrays}},
url = {https://www.nature.com/articles/s41586-023-06927-3},
volume = {626},
year = {2023}
}

@unpublished{Manetsch2024,
archivePrefix = {arXiv},
arxivId = {2403.12021},
author = {Manetsch, Hannah J. and Nomura, Gyohei and Bataille, Elie and Leung, Kon H. and Lv, Xudong and Endres, Manuel},
eprint = {2403.12021},
month = {mar},
title = {{A tweezer array with 6100 highly coherent atomic qubits}},
url = {http://arxiv.org/abs/2403.12021}
}

@article{Picard2024a,
archivePrefix = {arXiv},
arxivId = {2406.15345},
author = {Picard, Lewis R. B. and Park, Annie J. and Patenotte, Gabriel E. and Gebretsadkan, Samuel and Wellnitz, David and Rey, Ana Maria and Ni, Kang-Kuen},
doi = {10.1038/s41586-024-08177-3},
eprint = {2406.15345},
isbn = {4158602408},
issn = {0028-0836},
journal = {Nature (London)},
month = {jan},
number = {8047},
pages = {821--826},
publisher = {Springer US},
title = {{Entanglement and iSWAP gate between molecular qubits}},
url = {http://arxiv.org/abs/2406.15345 https://www.nature.com/articles/s41586-024-08177-3},
volume = {637},
year = {2025}
}

@article{Guttridge2023,
archivePrefix = {arXiv},
arxivId = {2303.06126},
author = {Guttridge, Alexander and Ruttley, Daniel K. and Baldock, Archie C. and Gonz{\'{a}}lez-F{\'{e}}rez, Rosario and Sadeghpour, H. R. and Adams, C. S. and Cornish, Simon L.},
doi = {10.1103/PhysRevLett.131.013401},
eprint = {2303.06126},
issn = {0031-9007},
journal = {Phys. Rev. Lett.},
month = {jul},
number = {1},
pages = {013401},
pmid = {37478436},
title = {{Observation of Rydberg Blockade Due to the Charge-Dipole Interaction between an Atom and a Polar Molecule}},
url = {https://link.aps.org/doi/10.1103/PhysRevLett.131.013401},
volume = {131},
year = {2023}
}

@article{Kaufman2021,
author = {Kaufman, Adam M. and Ni, Kang-Kuen},
doi = {10.1038/s41567-021-01357-2},
isbn = {4156702101357},
issn = {1745-2473},
journal = {Nat. Phys.},
month = {dec},
number = {12},
pages = {1324},
publisher = {Springer US},
title = {{Quantum science with optical tweezer arrays of ultracold atoms and molecules}},
url = {https://www.nature.com/articles/s41567-021-01357-2},
volume = {17},
year = {2021}
}

@article{Cornish2024,
archivePrefix = {arXiv},
arxivId = {2401.05086},
author = {Cornish, Simon L. and Tarbutt, Michael R. and Hazzard, Kaden R. A.},
doi = {10.1038/s41567-024-02453-9},
eprint = {2401.05086},
issn = {1745-2473},
journal = {Nat. Phys.},
month = {may},
number = {5},
pages = {730--740},
publisher = {Springer US},
title = {{Quantum computation and quantum simulation with ultracold molecules}},
url = {http://dx.doi.org/10.1038/s41567-024-02453-9 https://www.nature.com/articles/s41567-024-02453-9},
volume = {20},
year = {2024}
}

@article{DeMille2002,
author = {DeMille, D.},
doi = {10.1103/PhysRevLett.88.067901},
issn = {0031-9007},
journal = {Phys. Rev. Lett.},
month = {jan},
number = {6},
pages = {067901},
title = {{Quantum Computation with Trapped Polar Molecules}},
url = {https://link.aps.org/doi/10.1103/PhysRevLett.88.067901},
volume = {88},
year = {2002}
}

@article{Anderegg2019,
archivePrefix = {arXiv},
arxivId = {1902.00497},
author = {Anderegg, Lo{\"{i}}c and Cheuk, Lawrence W. and Bao, Yicheng and Burchesky, Sean and Ketterle, Wolfgang and Ni, Kang Kuen and Doyle, John M.},
doi = {10.1126/science.aax1265},
eprint = {1902.00497},
issn = {10959203},
journal = {Science},
number = {6458},
pages = {1156--1158},
pmid = {31515390},
title = {{An optical tweezer array of ultracold molecules}},
volume = {365},
year = {2019}
}

@article{Wang2022,
archivePrefix = {arXiv},
arxivId = {2204.05293},
author = {Wang, Kenneth and Williams, Conner P. and Picard, Lewis R.B. and Yao, Norman Y. and Ni, Kang-Kuen},
doi = {10.1103/PRXQuantum.3.030339},
eprint = {2204.05293},
issn = {2691-3399},
journal = {PRX Quantum},
month = {sep},
number = {3},
pages = {030339},
publisher = {American Physical Society},
title = {{Enriching the Quantum Toolbox of Ultracold Molecules with Rydberg Atoms}},
url = {https://doi.org/10.1103/PRXQuantum.3.030339 https://link.aps.org/doi/10.1103/PRXQuantum.3.030339},
volume = {3},
year = {2022}
}

@article{Picard2024,
archivePrefix = {arXiv},
arxivId = {2401.13659},
author = {Picard, Lewis R. B. and Patenotte, Gabriel E. and Park, Annie J. and Gebretsadkan, Samuel F. and Ni, Kang-Kuen},
doi = {10.1103/PRXQuantum.5.020344},
eprint = {2401.13659},
issn = {2691-3399},
journal = {PRX Quantum},
month = {may},
number = {2},
pages = {020344},
title = {{Site-Selective Preparation and Multistate Readout of Molecules in Optical Tweezers}},
url = {http://arxiv.org/abs/2401.13659 https://link.aps.org/doi/10.1103/PRXQuantum.5.020344},
volume = {5},
year = {2024}
}

@article{ARC2017,
title = {ARC: An open-source library for calculating properties of alkali Rydberg atoms},
journal = {Computer Physics Communications},
volume = {220},
pages = {319-331},
year = {2017},
issn = {0010-4655},
doi = {https://doi.org/10.1016/j.cpc.2017.06.015},
url = {https://www.sciencedirect.com/science/article/pii/S0010465517301972},
author = {N. Šibalić and J.D. Pritchard and C.S. Adams and K.J. Weatherill}}

@article{zhu2025probing,
  title = {Probing Dipolar Interactions between Rydberg Atoms and Ultracold Polar Molecules},
  author = {Zhu, Lingbang and Luke, Jeshurun and Shaham, Roy and Liu, Yi-Xiang and Ni, Kang-Kuen},
  journal = {Phys. Rev. Lett.},
  volume = {135},
  issue = {15},
  pages = {153001},
  numpages = {6},
  year = {2025},
  month = {Oct},
  publisher = {American Physical Society},
  doi = {10.1103/48rk-sxfs},
  url = {https://link.aps.org/doi/10.1103/48rk-sxfs}
}

@article{young2025data,
  title={Data-insensitive cooling of polar molecules with Rydberg atoms},
  author={Young, Jeremy T and Belyansky, Ron and Ni, Kang-Kuen and Gorshkov, Alexey V},
  journal={arXiv preprint arXiv:2507.10671},
  year={2025}
}

@article{kuznetsova2016rydberg,
  title={Rydberg-atom-mediated nondestructive readout of collective rotational states in polar-molecule arrays},
  author={Kuznetsova, Elena and Rittenhouse, Seth T and Sadeghpour, HR and Yelin, Susanne F},
  journal={Physical Review A},
  volume={94},
  number={3},
  pages={032325},
  year={2016},
  publisher={APS}
}

@article{PhysRevA.88.032709,
  title = {Long-range interactions between polar alkali-metal diatoms in external electric fields},
  author = {Lepers, M. and Vexiau, R. and Aymar, M. and Bouloufa-Maafa, N. and Dulieu, O.},
  journal = {Phys. Rev. A},
  volume = {88},
  issue = {3},
  pages = {032709},
  numpages = {8},
  year = {2013},
  month = {Sep},
  publisher = {American Physical Society},
  doi = {10.1103/PhysRevA.88.032709},
  url = {https://link.aps.org/doi/10.1103/PhysRevA.88.032709}
}

@phdthesis{wang2024dual,
  title={Dual Species Atom Arrays for Quantum Simulation and Computation},
  author={Wang, Kenneth},
  year={2024},
  school={Harvard University}
}

@article{zhang2026quantum,
  title={Quantum logic control and entanglement in hybrid atom-molecule arrays},
  author={Zhang, Chi and Murciano, Sara and Tantivasadakarn, Nathanan and Finkelstein, Ran},
  journal={arXiv preprint arXiv:2602.12909},
  year={2026}
}

@article{Young2026detection,
  title={Simultaneous nondestructive measurement of many polar molecules using Rydberg atoms},
  author={Young, Jeremy T and Ni, Kang-Kuen and Gorshkov, Alexey V},
  journal={arXiv preprint arXiv:2601.08921},
  year={2026}
}

@article{centralspinpaper,
  title = {Quantum simulation of the central spin model with a Rydberg atom and polar molecules in optical tweezers},
  author = {Dobrzyniecki, Jacek and Tomza, Micha\l{}},
  journal = {Phys. Rev. A},
  volume = {108},
  issue = {5},
  pages = {052618},
  numpages = {23},
  year = {2023},
  month = {Nov},
  publisher = {American Physical Society},
  doi = {10.1103/PhysRevA.108.052618},
  url = {https://link.aps.org/doi/10.1103/PhysRevA.108.052618}
}

@article{jandura2022time,
  title={Time-optimal two-and three-qubit gates for Rydberg atoms},
  author={Jandura, Sven and Pupillo, Guido},
  journal={Quantum},
  volume={6},
  pages={712},
  year={2022},
  publisher={Verein zur F{\"o}rderung des Open Access Publizierens in den Quantenwissenschaften}
}

@article{fu2022high,
  title={High-fidelity entanglement of neutral atoms via a Rydberg-mediated single-modulated-pulse controlled-phase gate},
  author={Fu, Zhuo and Xu, Peng and Sun, Yuan and Liu, Yang-Yang and He, Xiao-Dong and Li, Xiao and Liu, Min and Li, Run-Bing and Wang, Jin and Liu, Liang and others},
  journal={Physical Review A},
  volume={105},
  number={4},
  pages={042430},
  year={2022},
  publisher={APS}
}

@article{theis2016high,
  title={High-fidelity Rydberg-blockade entangling gate using shaped, analytic pulses},
  author={Theis, LS and Motzoi, F and Wilhelm, FK and Saffman, M},
  journal={Physical Review A},
  volume={94},
  number={3},
  pages={032306},
  year={2016},
  publisher={APS}
}

@article{echternach2001universal,
  title={Universal quantum gates for single cooper pair box based quantum computing},
  author={Echternach, P and Williams, Colin P and Dultz, SC and Delsing, P and Braunstein, SL and Dowling, JP},
  journal={Quantum Information and Computation},
  volume = "1",
    pages = "143-150",
    year = "2001"
}

@article{grassl1997codes,
  title={Codes for the quantum erasure channel},
  author={Grassl, Markus and Beth, Th and Pellizzari, Thomas},
  journal={Physical Review A},
  volume={56},
  number={1},
  pages={33},
  year={1997},
  publisher={APS}
}

@book{gottesman1997stabilizer,
  title={Stabilizer codes and quantum error correction},
  author={Gottesman, Daniel},
  year={1997},
  publisher={California Institute of Technology}
}

@article{wu2022erasure,
  title={Erasure conversion for fault-tolerant quantum computing in alkaline earth Rydberg atom arrays},
  author={Wu, Yue and Kolkowitz, Shimon and Puri, Shruti and Thompson, Jeff D},
  journal={Nature communications},
  volume={13},
  number={1},
  pages={4657},
  year={2022},
  publisher={Nature Publishing Group UK London}
}

@article{Yung2003AnEE,
  title={An exact effective two-qubit gate in a chain of three spins},
  author={Man-Hong Yung and Debbie W. Leung and Sougato Bose},
  journal={Quantum Inf. Comput.},
  year={2004},
  volume={4},
  pages={174-185},
  url={https://api.semanticscholar.org/CorpusID:33847284}
}

@article{ruttley2026harnessing,
  title={Harnessing resonant dipolar interactions in a hybrid atom-molecule quantum system},
  author={Ruttley, Daniel K and Hepworth, Tom R and Garc{\'\i}a-Garrido, Juan M and Rich, Caleb JH and Gonz{\'a}lez-F{\'e}rez, Rosario and Guttridge, Alexander and Cornish, Simon L},
  journal={arXiv preprint arXiv:2607.15976},
  year={2026}
}

@article{poyatos1997complete,
  title={Complete characterization of a quantum process: the two-bit quantum gate},
  author={Poyatos, JF and Cirac, J Ignacio and Zoller, Peter},
  journal={Physical Review Letters},
  volume={78},
  number={2},
  pages={390},
  year={1997},
  publisher={APS}
}



\let\oldref\ref
\newcommand{\refbrack}[1]{(\oldref{#1})}
\appendix
\mbox{~}

\section{\large{End Matter}}
\vspace{-.3cm}
\twocolumngrid

\emph{Appendix A: Dressing parameters}.---In this appendix we discuss the analytic solutions for the dressing parameters in the case of three dressed Rydberg levels.
The dressing equations are invariant under simultaneously transforming $\ket{b} \rightarrow-\ket{b}$ and $\Omega_\sigma \rightarrow -\Omega_\sigma$, while setting $\Delta_{\pi,\sigma} \rightarrow -\Delta_{\pi,\sigma} $ transforms the dressed-state energies as $E_{ab}\rightarrow - E_{ab}$.
All eight solutions can be accessed from a single solution by tuning the signs of $\alpha$ and $\Omega_\pi$. 
In this manuscript we therefore pick one of the solutions without loss of generality.
The dressed-state coefficients used in the main text are given by
\begin{align}
    \begin{split}
        &a_p=-\sqrt{1-\alpha^2}, \quad b_p=-\alpha, \\
        &a_\pi =\frac{\alpha}{\sqrt{1+2\mathcal{M}^2}}, \quad b_\pi=- \sqrt{\frac{1-\alpha^2}{1+2\mathcal{M}^2}}, \\
       & a_\sigma= -\frac{\mathcal{M} \alpha \sqrt{2}}{\sqrt{1+2\mathcal{M}^2}}, \quad b_\sigma=\mathcal{M} \sqrt{\frac{2-2 \alpha^2}{1+2\mathcal{M}^2}},\\
    \end{split}
\label{eqn:stateparams}
\end{align}
where $\mathcal{M}=\mu_\pi/\mu_\sigma$, and the free parameter $|\alpha|\leq1$ is related to the ratio between the detunings and the Rabi frequencies. Specifically, the associated drive parameters are
\begin{equation}
    \begin{split}
        \Omega_\sigma &=-\sqrt{2} \mathcal{M} \Omega_\pi,\\
        \Delta_\pi &=\sqrt{1+2 \mathcal{M}^2} \frac{2 \alpha^2 -1}{\alpha \sqrt{1-\alpha^2}} \Omega_\pi, \quad\Delta_\sigma =  \Delta_\pi,
    \end{split}
\label{eqn:driveparams}
\end{equation}
where the negative Rabi frequency $\Omega_\sigma$ corresponds to a $180^\circ$ phase shift between the $\pi$ and $\sigma_+$ drives.  Although the two detunings $\Delta_\pi,\Delta_\sigma$ are equal in this three-state scheme, when more states are involved, the required detunings generally differ.
Note that, at $|\alpha|=0,1$, the detunings diverge and the dressing breaks down. The parameter regime around these values will therefore be avoided.
The Rabi frequencies $\Omega_{\pi,\sigma}$ must be significantly stronger than both the dipolar and vdW interactions between atoms for the dressed-state picture to be valid but cannot be so large as to become comparable to the energy splitting between additional neighboring Rydberg states, which would result in a cascade of many more states becoming dressed.

\emph{Appendix B: Dressing of additional Rydberg states}.---In this appendix we discuss how to nullify the dipolar Rydberg-Rydberg interactions when more than three states are dressed by the microwave fields.
When using Rabi frequencies comparable to the fine-structure splitting or when the drive polarization is impure, additional Rydberg states are dressed.
This modifies the dressing scheme and makes the analytical solution given in Eqs.~\refbrack{eqn:stateparams} and~\refbrack{eqn:driveparams} invalid. Nonetheless it is still possible to find a numerical solution which minimizes the dipolar Rydberg-Rydberg interactions. 
Focusing on the case in which the drive strength is comparable to the fine-structure splitting, for each of the $\ket{p}, \ket{d}$ states, one additional state differing only in $J$ is dressed each.
No additional $\ket{s}$ state which only differs in $J$ exists because the electron spin of Cs is $S=1/2$.  
We denote the two extra states by $\ket{p'},\ket{d'}$.
Taking these two states into account, the modified dressed states are given by
\begin{equation}
    \begin{split}
        \ket{a'} & =  a_p \ket{p} + a_\pi \ket{d} + a_\sigma \ket{s}+ a_p' \ket{p'}+ a_\pi' \ket{d'} ,\\
        \ket{b'} & = b_p \ket{p} + b_\pi \ket{d} + b_\sigma \ket{s} + b_p' \ket{p'}+b_\pi' \ket{d'} .
    \end{split}
\end{equation}
The associated microwave dressing Hamiltonian is
\begin{subequations}
\begin{equation}
    H_{\text{mw}}' = \left( 
    \begin{array}{ccccc}
        0 & 0 & \Omega_\pi & \Omega_\sigma & f_\pi \Omega_\pi  \\
        0 & \epsilon_{p'} & g_\pi \Omega_\pi & g_\sigma \Omega_\sigma & g_\pi' \Omega_\pi \\
        \Omega_\pi & g_\pi \Omega_\pi & - \Delta_\pi & 0 & 0\\
        \Omega_\sigma &  g_\sigma \Omega_\sigma &  0 & - \Delta_\sigma & 0\\
        f_\pi \Omega_\pi & g_\pi' \Omega_\pi & 0 & 0 & -\Delta_\pi+\epsilon_{d'}\\
    \end{array}
    \right),
\end{equation}
\begin{equation}
    f_\pi = \frac{\langle d'| d_{0}^R| p \rangle}{\langle d| d_{0}^R| p \rangle},
    \quad
       g_\pi = \frac{\langle d|  d_{0}^R| p' \rangle}{\langle d| d_0^R| p \rangle},
       \quad
        g_\sigma = \frac{\langle p'| d_{+}^R| s \rangle}{\langle p| d_+^R| s \rangle},
\end{equation}
\begin{equation}
    g_\pi' = \frac{\langle d'| d_0^R| p' \rangle}{\langle d| d_0^R| p \rangle}, 
    \quad
     \epsilon_{p'} = E_{p'}-E_{p},
    \quad
     \epsilon_{d'} = E_{d'}-E_{d}.
\end{equation}

\end{subequations}
The requirements for nullifying the dipole-dipole interactions are
\begin{subequations}
\begin{equation}
\begin{split}
   & V_{aa} \propto  ( a_p \left(a_\pi +f_\pi a_\pi' \right) +a_p' \left(g_\pi a_\pi +g_\pi' a_\pi' \right) ) ^2 \mu_\pi^2\\
   &-( a_\sigma \left(a_p+ g_\sigma a_p' \right) )^2 \mu_\sigma^2 /2 =0\\
\end{split}
\end{equation} 
\begin{equation}
    \begin{split}
         &V_{ab} \propto \left(a_\pi \left(a_p+ g_\pi a_p' \right) + a_\pi'\left(f_\pi a_p + g_\pi' a_p' \right) \right) \times \\
         &\left(b_\pi \left(b_p+ g_\pi b_p' \right) + b_\pi'\left(f_\pi b_p + g_\pi' b_p' \right) \right) \mu_\pi^2\\
         &- a_\sigma b_\sigma\left(a_p +g _\sigma a_p' \right) \left(b_p +g _\sigma b_p' \right)\mu_\sigma^2/2=0.
    \end{split}
\end{equation}
\end{subequations}
Note that due to the form of the five-level $V_{bb}$ and $\tilde{V}_{ab}$ interactions, similarly to the three-level case, these will also be nullified if $V_{aa},V_{ab}$ are.

As mentioned before, the involvement of more states generally alters the dressing parameter solutions. The deviation from the original three-state solution depends on, among other things, how close in energy the additional state is to the other addressed states and the strength of the drive.
As in the three-state case, the solutions are not unique; two notable branches of solutions are those that are mostly made up of the original $\ket{s}, \ket{p}, \ket{d}$ states and the solutions that mostly consist of the additional states  $\ket{s}, \ket{p'}, \ket{d'}$. 
Generally the former are the more desirable solutions, as the original three states should have been chosen specifically to maximize the Rydberg-molecule interaction strength given in Eq.~(\ref{eqn:V_rydmol}) and have an energy splitting close to that of the molecule.
We therefore find solutions to the full five-state problem by using the three-state parameters at low Rabi frequencies and then adiabatically increase $\Omega_\pi$ to find a solution at the desired $\Omega_\pi$ which is connected to the original three-state solution. 
The calculations shown in Fig.~\ref{fig:ryd-mol_resonance} take into account the five-level corrections.

\emph{Appendix C: Analysis of gate infidelity caused by motion and Rydberg decay.}---
In this appendix we outline our treatment of the motion of the atom and molecules, the decay of the dressed atom, and the resulting gate infidelity.
Due to the large size of the required Hilbert space, it is computationally demanding to simulate the dynamics of two molecules and one atom in optical tweezers in three dimensions.
We therefore treat the motion along the three axes separately by performing a single one-dimensional simulation for motion along each axis to obtain an estimate of the gate fidelity.
The trapping potentials of the molecules are approximated as harmonic oscillators and are assumed to be state-independent.
Before the gate, the atom is assumed to be trapped with trapping frequency $\omega_R^\gamma$, but it is untrapped during the gate. 
We expand the position and momentum operators of the free atom in the harmonic-oscillator basis of the initial trap.
Setting $\hbar=1$, the corresponding Hamiltonian is given by
\begin{equation}
    H^0_\gamma=(p_R^\gamma)^2/(2 m_R)+\sum_{i=1,2} \omega^\gamma_i \left(a^\dagger_i a_i +1/2\right),
\end{equation}
 where $1,2$ label the two molecules, $R$ labels the Rydberg atom, $m_j$ is the mass of particle $j$, $p^\gamma_j=i\sqrt{ m_j\omega_j^\gamma/2} \left(a_j^\dagger -a_j\right)$ is the momentum operator, $\omega^\gamma_j$ is the trapping frequency for particle $j$ along direction $\gamma$ and $\gamma \in (x,y,z)$.
When simulating gates with a trapped atom, we use the Hamiltonian $\tilde{H}^0_\gamma=\sum_{i=1,2,R} \omega^\gamma_i ( a^\dagger_i a_i+1/2)$.
The interaction Hamiltonians for simulations along the $x,y,z$ directions are given by
\begin{equation}
    \begin{split}
        H^{\rm{int}}_x&=\sum_{j=1,2} \frac{C_J}{(r_0+(-1)^j (r_j^x-r_R^x))^3} \left(\ketbra{1_j a}{0_j b} +H.c. \right), \\
        H^{\rm{int}}_y&=\sum_{j=1,2} \frac{C_J}{\left(r_0^2+\left(r_j^y-r_R^y \right)^2 \right)^{3/2}} \left(\ketbra{1_j a}{0_j b} +H.c. \right),\\
        H^{\rm{int}}_z&=\sum_{j=1,2} \frac{C_J \left(1-3 \left(r_j^z-r_R^z \right)^2 /\left(r_0^2+\left(r_j^z-r_R^z \right)^2 \right) \right)}{\left(r_0^2+\left(r_j^z-r_R^z \right)^2 \right)^{3/2}} \\
         &\times \left(\ketbra{1_j a}{0_j b} +H.c. \right),\\
    \end{split}
\end{equation}
where $C_J$ is the Rydberg-molecule interaction coefficient, the position operator along axis $\gamma$ is given by $r^\gamma_j=\sqrt{\hbar/ 2 m_j \omega^\gamma_j} (a_j+a^\dagger_j)$, and $(r_j^z-r_R^z )^2 /(r_0^2+(r_j^z-r_R^z )^2 )$ corresponds to $\cos{\left(\delta\theta_{iR} +\pi/2 \right)}^2$, where $\delta \theta_{iR}$ is the deviation of $\theta_{iR}$ from the intended value of $\pi/2$.
The total Hamiltonian is given by $H^{\rm tot}_\gamma=H^{0}_\gamma+H^{\rm int}_\gamma$.

We assume the $z$-axis to be the axial direction (along the tweezer k-vector) and hence the weakly trapped direction.
The masses of the two molecules and the Rydberg atom are $m_{1,2}$ and $m_R$, respectively, and the trapping frequencies are assumed to be equal for the atom and molecules, $\omega_{x,y}=2 \pi \times 300 $ kHz, $\omega_z=2 \pi \times 60$ kHz.
The simulations are performed by numerically propagating the initial density matrix $\rho(0)$, where the particles are initialized in a thermal state of motion with excitation number $\bar{n}_x$ along $x$ and in the ground state along $y$ and $z$. 
\\

We estimate the errors caused by the decay of the dressed Rydberg atom by simulating the gate dynamics with an effective loss model described by the Lindblad master equation
\begin{equation}
   \dot \rho = -i \left[ H_G, \rho \right] + \sum_{j=a,b}\left(\ell_j\rho  \ell_j^\dagger - \frac12\lbrace \ell_j^\dagger\ell_j,\rho\rbrace \right),
   \label{eqn:master}
\end{equation}
where $H_G$ is given by Eq.~(\ref{eqn:H_G}) and $\ell_j$ is the jump operator accounting for decay from state $\ket{j}$.
We assume that the $\ket{a}$ and $\ket{b}$ states irreversibly decay to a set of other states which do not participate in the mediated gate and are collectively labeled as the ``bath'' state $\ket{\mathcal{B}}$.
This model assumes that an atom which decays to $\ket{\mathcal{B}}$ by spontaneous emission or is incoherently pumped to $\ket{\mathcal{B}}$ by black-body radiation cannot return to the $\ket{a},\ket{b}$ manifold on the timescale of a gate.
If repopulation of the $\ket{a,b}$ states does occur, this can result in unflagged errors in the post-selection or error-correction step.
The associated jump operators are given by
\begin{equation}
    \ell_{a}=\sqrt{\Gamma_a}\ketbra{\mathcal{B}}{a}, ~~\ell_{b}=\sqrt{\Gamma_b}\ketbra{\mathcal{B}}{b},
\end{equation}
where $\Gamma_{a}$ and $\Gamma_b$ are the effective decay rates of states $\ket{a,b}$ respectively.
\\

The average gate fidelities are characterized by initializing the atom in $\ket{a}$ and the molecules in one of the 16 tomographically complete two-molecule states constructed by tensor products of $\ket{0},\ket{1},\frac{1}{\sqrt{2}}\left( \ket{0}+  \ket{1}\right)$ and $\frac{1}{\sqrt{2}}\left( \ket{0}+i \ket{1}\right)$~\cite{poyatos1997complete}. The gate is subsequently applied for time $t_G$. 
Next, the internal states of the atom and the motion of all particles are traced out from $\rho(t_G)$ and we use the resulting two-molecule density matrix $\rho^{1,2}(t_G)$ to compute the average fidelity by reconstructing the transfer operators $R_{i,j}$ according to the procedure outlined in~\cite{poyatos1997complete}.

In order to calculate the success-weighted fidelity after post-selection, we project onto the Rydberg $\ket{a}$ state to obtain $\rho'(t_G)= \ketbra{a}{a}\rho (t_G) \ketbra{a}{a}$, which is used to reconstruct the trace-nonincreasing transfer operators $\tilde{R}_{i,j}$ and calculate the corresponding unnormalized average success-weighted fidelity $\bar{\mathcal{F}}_a$, averaged over the 16 input states.
The total post-selection success rate is characterized by $\mathcal{P}_s=\Tr{\rho(t_G) \ketbra{a}{a}}$ and is also approximated as the sum of the $x,y,z-$motion and decay contributions and subsequently averaged over the 16 two-molecule input states. 
Finally, the success weighted post-selected fidelity is given by $\bar{\mathcal{F}}_{\rm {ps}}=\bar{\mathcal{F}}_a/\mathcal{\bar{P}}_s$.

\begin{figure}
    \centering
    \includegraphics[scale=1]{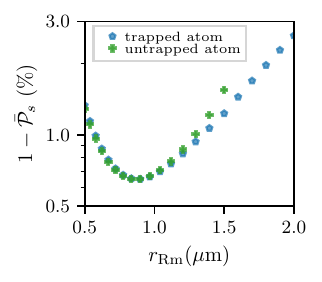}
    \caption{
   Average post-selection rejection probability $1-\bar{\mathcal{P}}_s$ as a function of the atom-molecule spacing $r_{\rm Rm}$.
   The total probability is a leading-order estimate computed by summing the contributions of motion along the $x,y,z$ axes and decay of the dressed Rydberg atom. The probabilities are averaged over 16 two-molecule input states.
   The untrapped atom data corresponds to calculations for an atom which is released from its trap at the start of the gate.}
    \label{fig:failrate}
\end{figure}

From the curves shown in Fig.~\ref{fig:failrate}, it is apparent that the post-selection rejection rate is higher for gates involving an untrapped atom due to the expansion of the atom's wave packet.

\emph{Appendix D: Molecular hyperfine structure}.--- In this appendix, we discuss the role of the molecular hyperfine structure.
As discussed in~\cite{Young2026detection}, the tensor interaction between the nuclear dipole moments, the quadrupole interaction and the spin-rotation coupling in the hyperfine Hamiltonian of bi-alkali molecules~\cite{Aldegunde2017} can couple molecular states with different projections of the rotational angular momentum $m_N$.
These couplings can be made off-resonant using the state-dependent light shifts caused by the microwave dressing discussed in the main text. However, as the considered dressing is off-resonant on the molecules and the Rabi frequencies are limited by the Rabi frequency on the Rydberg atoms, the resulting light shifts are not guaranteed to be sufficient.
If the light shifts induced by the three-level dressing are insufficient, an additional $\pi$-polarized microwave field, resonant with the $N=1 \leftrightarrow N=2$ transition in the molecules can be applied. 
As the Rabi frequency of this dressing field is not limited by the Rydberg atom, large light shifts can be applied, while the Clebsch-Gordan coefficients ensure that $\ket{N=1,~m_N=0}$ sees a different light shift than the other $\ket{N=1}$ states.
When this additional dressing is applied, dipolar interactions couple $\ket{N=0,m_N=0}$ to the superposition $\left(\ket{N=1,m_N=0}+\ket{N=2,m_N=0} \right)/\sqrt{2}$ and as a result, the Rydberg-molecule interaction is reduced by a factor $1/\sqrt{2}$.

\emph{Appendix E: Unused third dressed Rydberg state}.---In this appendix we discuss the role of the unused third dressed state in the three-level dressing scheme.
In the mediation scheme discussed in this manuscript, only two of the three dressed Rydberg states have been utilized.
Due to the degeneracy of the required microwave detunings, the state coefficients of the third state are independent of $\Omega_\pi$ and $\alpha$. 
Labeling this state as $\ket{c}=c_p \ket{p}+ c_\pi \ket{d} +c_\sigma \ket{s}$, the state coefficients are given by $c_p=0,c_\pi=\frac{\mathcal{M}\sqrt{2}}{\sqrt{1+2\mathcal{M}^2}},c_\sigma=\frac{1}{\sqrt{1+2\mathcal{M}^2}}$.
Although diagonal interactions involving $\ket{c}$ are also nullified by the microwave dressing, flip-flop interactions involving $\ket{c}$ are not guaranteed to be zero.
Therefore, due to the presence of dipolar flip-flop interactions, the $\ket{c}$ state is not suitable to be used to mediate gates between molecules, as these interactions will result in delocalized Rydberg excitations, which we wish to avoid using the microwave dressing scheme.
\end{document}